\documentclass{article}

\usepackage{amsmath,amssymb,amsfonts}

\usepackage{graphicx}

\usepackage{listings}

\newtheorem{rule-of-thumb}[theorem]{Definition} 

\begin{document}
\title{Airplane Orbits and Satellite Orbits and Orbitfall: Physics Hidden in Plain Sight}

\author{John P. Boyd \\
Department of Climate \& Space
Sciences and Engineering  \\University of
Michigan, 2455 Hayward Avenue, Ann Arbor MI 48109 \\
jpboyd@umich.edu}

\maketitle

To be submitted to \emph{American Journal of Physics}

\begin{abstract} 
An airplane flying at constant speed and altitude is an example of physics invisible to  the pilots and passengers, but visible to remote observers and manifest in the mathematics. The optimum flight path is an arc of a  Great Circle, specifically that circle which is the result of rotating the equator to intersect the origin and destination airports. In order that the velocity vector remain parallel to the surface of the spherical earth, a centripetal force is required to rotate the velocity without altering its magnitude. This force must be radially inward and thus  parallel to the local vertical. The assertion that ``lift balances gravity" is only  an approximation. To follow the curve of the earth, the vertical component of aerodynamic lift must
be \emph{slightly weaker} than gravity  so that the plane can be in ``orbitfall".
  That is, to follow the curvature of the earth, maintaining a constant distance $R$ from the center of 
the earth while flying at a constant speed $V$ and generating a vertical lift per unit mass $L$, the plane and all inside it must fall towards the center of 
the earth with an acceleration, the ``orbitfall acceleration",  $g_{orbitfall} \equiv g-L=V^{2} / R$ 
where $g$ is the usual gravitational acceleration constant. 
If the plane travels a fraction $\digamma$ of the earth's circumference , then it must execute $\digamma$ of 
an outside loop. This requires that the pitch angle $\xi$ must rotate nose-down at a rate of $d \xi / dt=V/R$.
 The dynamic stability of a non-military aircraft makes this pitch change continuously without pilot 
intervention: the horizontal stabilizers act as weathervanes, suppressing small perturbations so 
that the longitudinal axis remains at a constant angle of attack. Neither pilot nor passenger is aware 
of the orbitfall or the automatic pitch changes --- physics invisible but essential.

\end{abstract}

\bigskip

\begin{quotation}
``All that's needed for a plane to fly around a spherical 
earth is a miniscule difference betweeen lift and weight."
\end{quotation}
\hspace*{0.5in} --- ``roohif" in \cite{roohifAirplaneOrbits}

\bigskip

\section{Introduction}

\begin{figure}[h]
\centerline{\includegraphics[scale=0.35]{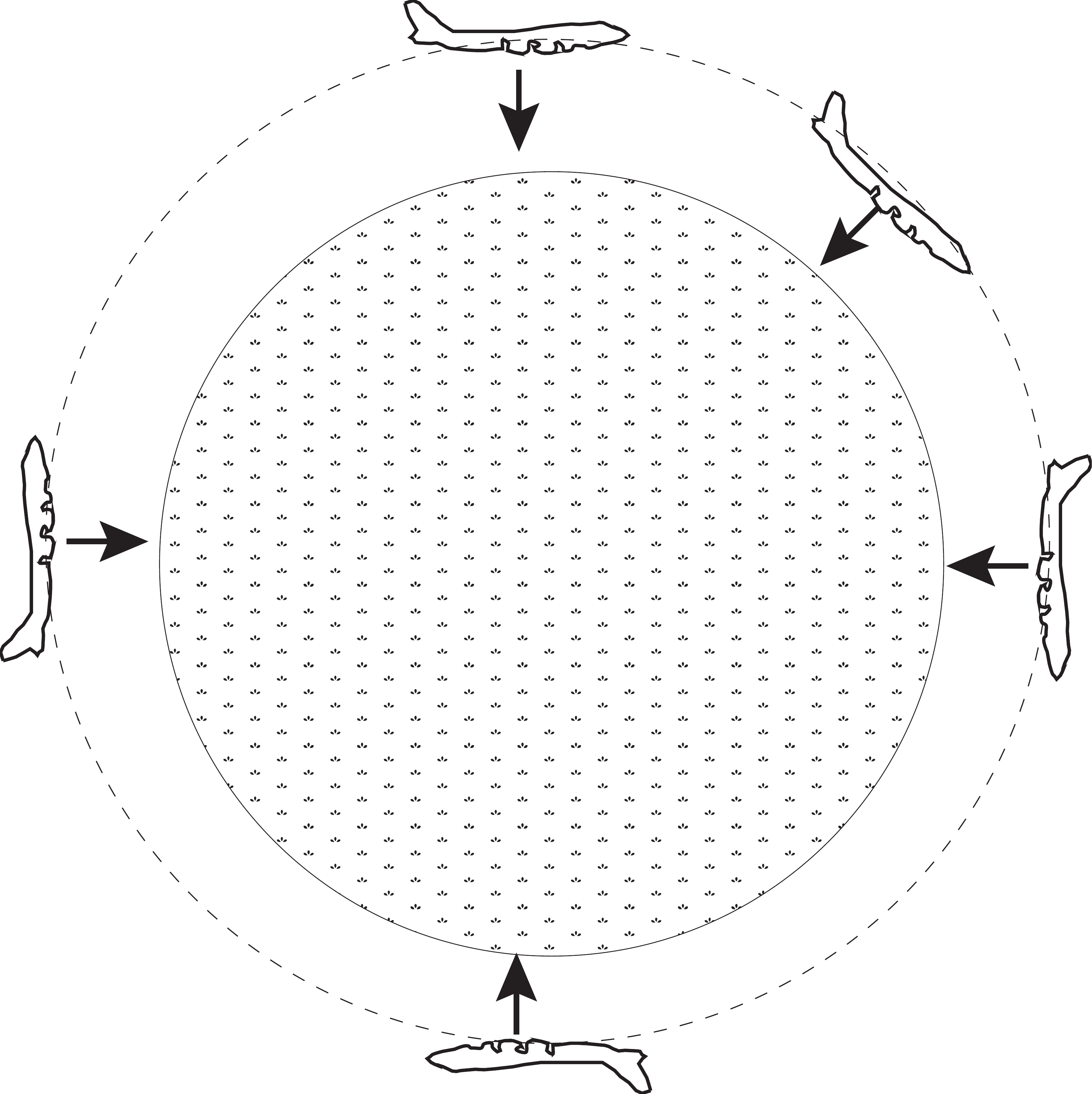}}
\caption{A schematic diagram of the flight of an airplane on an arc of a Great Circle.  The shaded disc is a 
side view of the sphere, which has a radius of roughly $R_{E} = 6.366 \times 10^{6}$ m. The dashed curve is the 
Great Circle which is the route of the aircraft; its radius is $R=R_{E} + H$ where $H$ is the altitude of
the jet, typically $H \approx 12$ km. The aircraft icons show the orientation of the aircraft at different 
points on the Great Circle. To fly a path of 1/8 or 1/4 or 1/2 of the Great Circle, the plane must pitch nose-down by 45 degrees, 90 degrees or 180 degees, respectively. These changes in the pitch angle $\xi$ are changes in this angle as seen by an observer off the globe, such as in a spaceship; this is the viewpoint of this schematic. In contrast,  the pilots and passengers in the plane perceive no change in pitch whatsoever. A centripetal acceleration is needed to rotate the 
velocity vector through an angle   which is equal to the change in the pitch angle $\xi$; this downward acceleration  has the magnitude 
$a_{orbitfall}= g - L$. The black arrows, attached to each aircraft icon, show the direction of the  orbitfall 
acceleration. This is always normal to the plane's velocity vector, that is,  radially inward (in spherical coordinates)  towards the center of the earth; this direction is downward in the local vertical direction as perceived by those on the airplane.     }
\label{FigOrbitfall_ill}
\end{figure}

If an aircraft is flying parallel to the earth's surface, maintaining a constant altitude and thus flying an arc of a Great Circle , then constant speed $V$ requires that  the net horizontal force (thrust of engines minus aerodynamic drag) must be zero. However, as shown 
in Fig.~\ref{FigOrbitfall_ill}, following the curve of the earth requires that the the pitch angle $xi$, the vertical and horizontal coordinates axes,  and the orientation of the velocity vector must \emph{rotate} in synchronization. For a 
flight that covers a fraction $\digamma$ of the circumference of the earth, all these angles and vectors must 
rotate through a total of $2 \pi \digamma$ radians.  

The velocity vector will rotate if and only if there is a centripetal acceleration,  a force acting towards 
the center of the Great Circle. 
 For movement at constant speed in a circle, the centripetal force is always a \emph{normal} force, that is, a  force  that acts perpendicular to the velocity vector. Normal forces \emph{turn} the velocity vector without altering the speed $V$. But what is the source of the centripetal force? And how do we calculate it as a function of $g$, 
$R$ and $V$ where $g$ is the gravitational constant at distance $R$ from the center of the planet,
\begin{equation}~\label{Eqg}
g = g_{E} \dfrac{R_{E}^{2}}{R^{2}}, \qquad g_{E}
\end{equation}where the gravitational constant $g_{E}$ at the earth's surface is about $9.81$ meters/second-squared.

Although space is three-dimensional, the Great Circle lies entirely within a plane that intersects the center of 
the planet. This allows us to use a two-dimensional polar coordinate system with the origin at the 
center of the earth, $r$ as the radial coordinate and $\theta$ as the polar angle. Newton's equations 
 in an inertial reference frame are, with the forces per unit mass on the right side (Eq. 2.14) of \cite{StommelMoore89})
\begin{eqnarray}
 \dfrac{d^{2} r}{dt^{2}} - r \left\{  \dfrac{d \theta}{dt}   \right\}^{2} & = &  - g \, + L \\
 \dfrac{d^{2} \theta}{dt^{2}} + 2 r \dfrac{dr}{dt} \dfrac{d \theta}{dt} & =  & 0 
\end{eqnarray}
where the speed $V=  r \, d \theta/dt$.footnote{Airplanes fly slightly nose up so as to generate a positive angle-of-attack, as explained in aeronautics texts; the lift vector is not purely vertical, but includes a small horizontal component that is cancelled by the thrust generated by the propeller or jet engine. Only the vertical component of lift is relevant to the vertical balance of forces.} 

Since for a constant altitude, constant velocity orbit, $r(t)=R$, $dr/dt=0$ and $r \, d theta/dt=V$, the equations simplify to
\begin{eqnarray}
  \dfrac{V^{2}}{R}  & = &   g \, - \,  L \\
  0 & = & 0
\end{eqnarray}

A space station or satellite is the special case that the lift is zero. The speed $V_{freefall}$ required for 
a space station to orbit without lift is
\begin{eqnarray}
  V_{freefall} & = &  \sqrt{ g \, R}
\end{eqnarray}
Every part of the space station including the astronauts is accelerating downward at an acceleration
 $g$ where $g$ is defined by Eq.~\ref{Eqg}. All these downward accelerating objects are said to be 
in ``freefall".  The astronauts 
and everything loose appear to be experiencing no gravity. This is an illusion, a spectacular counter-example to the maxim ``trust your senses"; in reality, everything in the space station is accelerating towards the center of the earth at an acceleration  $g$.
 The space station or satellite does not hit the surface of the earth, or even change distance from the surface because (i) the spacecraft has a high horizontal velocity and (ii) the surface of the earth is curving downward, away from the falling spacecraft.

An aircraft circling the earth at constant altitude is doing a perfect imitation of  
the space station except that the downward acceleration, the ``orbitallfall", is 
\begin{eqnarray}~\label{Eqa}
a_{orbitfall} = g - L \\
   & = & V^{2} / R
\end{eqnarray}

When the aerodynamic lift is not zero, it is possible to maintain constant altitude with much smaller 
speed $V$,
\begin{eqnarray}
V = V_{freefall} \, \sqrt{ 1 - L/g}
\end{eqnarray}

A numerical example is instructive.
 If the jet is traveling at 255 m/s, then $R=6.378 \times 10^{7}$ meters gives
 \begin{eqnarray}
V / V_{freefall} \approx 0.032 \qquad \Leftrightarrow \qquad 
\end{eqnarray}
or in other words the orbital speed of a space station is about thirty-one times larger
Eq.~\ref{Eqa} shows that the vertical lift is smaller than $g$ by slightly more than 1 part in a thousand or about 1 centimeter per second-squared.

Airplanes do not have instruments to measure lift directly; a pilot climbs to the target height and then 
adjusts the trim, pitch and thrust until the vertical velocity is zero. Physics then requires that the lift 
be slightly smaller than gravity.

\clearpage

\section{The Ever-Changing Pitch of the Airplane and Inherent Stability}

\begin{quote}
``A wing with a conventional aerofoil profile makes a negative contribution to longitudinal stability. This means that any disturbance (such as a gust) which raises the nose produces a nose-up pitching moment which tends to raise the nose further. With the same disturbance, the presence of a tailplane produces a restoring nose-down pitching moment, which may counteract the natural instability of the wing and make the aircraft longitudinally stable (in much the same way a weather vane always points into the wind). "
\end{quote}
\hspace*{0.5in} --- Wikipedia article \emph{Tailplane}

\bigskip

A bicycle is dynamically unstable if the wheels are not turning; stop pedaling, and
it will slow and then fall over in  seconds, which is why a kickstand is mandatory. 
When moving at sufficient speed, however, the wheels acts as gyroscopes, resisting perturbations,  so a child can shout, ``Look, Mom! No hands!"

 Airplanes in flight are also dynamically stable.  Without 
pilot intervention,  
aircraft automatically correct for small perturbations, such as  internal gravity waves, wind shear,  convective plumes and the constant buffeting by convective bubbles.

The horizontal parts of the tail fin are the ``horizontal stabilizers". The elevators, which allow manual control of pitch, are moveable panels embedded in the horizontal stabilizer.

\begin{figure}[h]
\centerline{\includegraphics[scale=0.35]{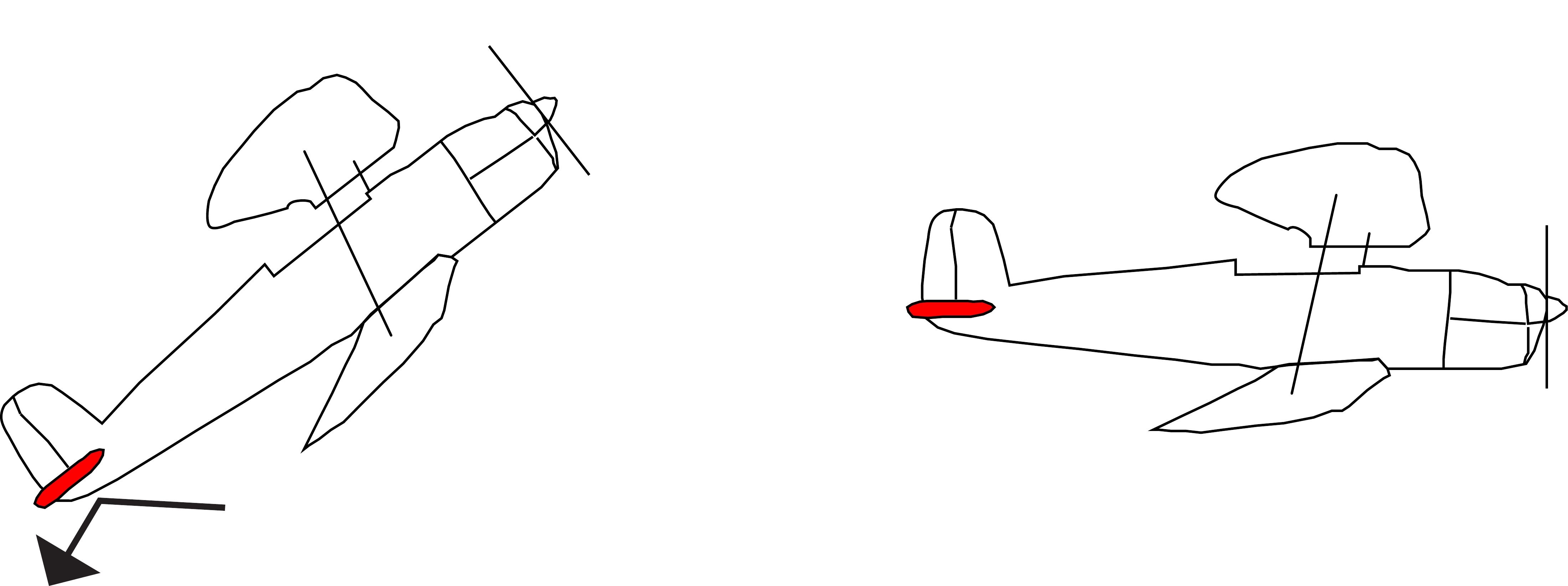}}
\caption{If the aircraft is pitched nose-high as shown on the left while the center-of-mass is moving horizontally, the airflow [black arrow] deflecting off  the horizontal stabilizer fin [red] will create a torque that will rotate the airplane nose-down.   This ``wind vane effect" realigns  pitch angle  as shown on the right.}.
\label{FigPitch_Horizontal_Stabilizer_Two_Plot}
\end{figure}

 The horizontal stabilizers provide stability in pitch. As the aircraft orbits around the earth at a constant altitude, the horizontal fins and the fuselage act like a weather vane; the flow deflecting off the the horizontal stabillizers  provides a torque that rotates the airplane about the pitch axis.  This axis is parallel to the line from one wingtip to the other wingtip. This dynamical stability, this weathervane effect, keeps the aircraft always at the proper pitch parallel to the sphere as  illustrated in  Fig.~\ref{FigPitch_Horizontal_Stabilizer_Two_Plot}.

\emph{Large} perturbations, of course, require manual adjustments by the pilot. Only \emph{small} perturbations are damped out by the  horizontal fins in the tail.

This is not a problem for constant altitude flight. As the airplane flies an arc of a circle, the pitch needs to be adjusted \emph{smoothly} and \emph{continuously} in what, from a calculus viewpoint, can be interpreted as an infinite number of infinitesimal corrections.

\section{Summary}

Proponents of the ``Flat Earth" have invoked airplane orbits as important evidence for their cause. 
Their argument notes, correctly, that (i) the aircraft must fall around the curve of the sphere and (iii) 
 the pitch of the jet must rotate through a fraction $\digamma$ of a full circle when flying an arc that is 
a fraction 
$\digamma$ of  the circumference of the earth.

  Their false part of their argument is the assertion that neither can happen without constant intervention of the pilots and awareness of pitch-down-and-fall by  the passengers; therefore, the absence of these perceptions implies that the surface of the planet is flat. We have shown that this not true; orbitalfall (downward acceleration) and pitch adjustments are automatic, unperceived by the passengers, requiring no action by the pilots. 

The other insight  is that satellite orbits are just a special case, namely the special case of \emph{zero  
lift}, of the larger family of objects flying at constant speed and altitude above a spherical planet.




\end{document}